\documentclass[12pt]{iopart}
\usepackage{iopams}
\usepackage{epsfig}

\begin{document}

\title[]{Shear Viscosity to Entropy within a Parton Cascade }

\author{A.El$^1$, C. Greiner$^1$ , Z. Xu$^1$}
\address{$^{1}$ Institut f\"ur Theoretische Physik,
 J.W. Goethe-Universit\"at Frankfurt am Main, Germany}

\begin{abstract}
The shear viscosity is calculated  
by means of the perturbative kinetic partonic cascade BAMPS with CGC initial conditons
for various saturation momentum scale $Q_s$. $\eta /s \approx $ 0.15 stays approximately
constant when going from RHIC to LHC.
 \end{abstract}

The measured momentum anisotropy parameter $v_2$
at RHIC energy can be well understood 
if the expanding quark-gluon matter is assumed to be described by ideal hydrodynamics.
This suggests that a strongly interacting and locally thermalized state of matter 
has been created which behaves almost like a perfect fluid. 
Since the initial situation of the quark-gluon system is far from thermal equilibrium, it is 
important to understand how and which microscopic partonic interactions can 
thermalize the system within a short timescale and can be responsible
as well for its (nearly) ideal hydrodynamical behaviour.
Furthermore one would like to know  the transport properties of the QGP,
most prominently the shear viscosity. 

A kinetic parton cascade (BAMPS) \cite{Xu:2004mz, Xu:2007aa} has been developed with 
strictly perturbative QCD inspired processes including for the first time inelastic (''Bremsstrahlung'') collisions $gg \leftrightarrow ggg $. The multiparticle back reation channel is treated fully consistently by respecting detailed balance within the same 
algorithm. In \cite{Xu:2007aa} it is demonstrated that the inelastic processes 
dominate the total transport collision rate and thus contribute much stronger 
to momentum isotropization then elastic ones. Within a default setting
of minijet initial conditions, the 
overall build up of elliptic flow $v_2$ can be reasonably described \cite{Xu:2005wv}
(a more dedicated study is presently undertaken \cite{Xu-next}).

One can thus expect to see thermalization of a QGP on a short time scale less than 1 $fm/c$ 
for LHC relevant initial conditions as can be seen in the evolution in time of 
the temperature and the momentum isotropy depicted in Fig.1. 
We apply Bjorken expanding geometry in one dimension. 
For the initial condition a simple Color Glass Condensate (CGC) 
gluon distribution is assumed: 
The initial partons are described by the boost-invariant 
form of the distribution function
$f(x,p)|_{z=0}=\frac{c}{\alpha_s \, N_c}\, \frac{1}{\tau_f}\, \delta(p_z)\, \Theta(Q_s^2-p_T^2)$
at a characteristic time $\tau_0 =c/(\alpha_s N_c Q_s)$.

Due to $3\rightarrow 2$ collisions the particle number first decreases (see Fig. 1) 
\cite{El:2006xj}.
This is in contrast to the idealistic ''Bottom-Up'' scenario of thermalization,
where an ongoing particle production in the soft sector ($p_T < \alpha_s Q_S$) is predicted
with a strong increase in the total particle number. The present calculation show that
the particle number roughly stays constant. For the above simple CGC 
parametrization
$Q_s=2$ GeV corresponds to RHIC energy whereas $Q_s \approx 3-4$ GeV is expected for LHC.

\begin{figure}
\hskip -0.5cm
\begin{minipage}[]{50mm}
\epsfxsize=5.0 cm
\epsfysize=5.0 cm
\epsfbox{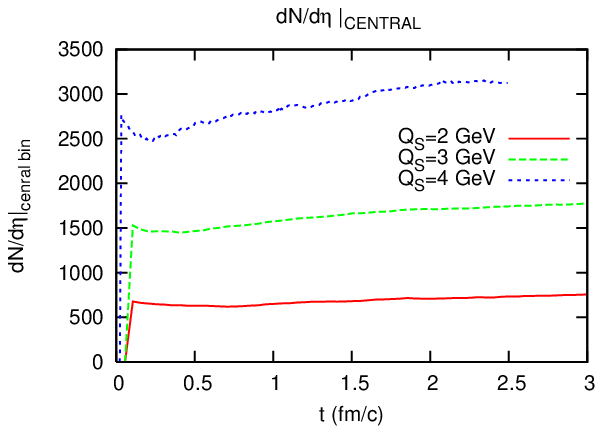}
\end{minipage}
\begin{minipage}[]{50mm}
\epsfxsize=5.0 cm
\epsfysize=5.0 cm
\epsfbox{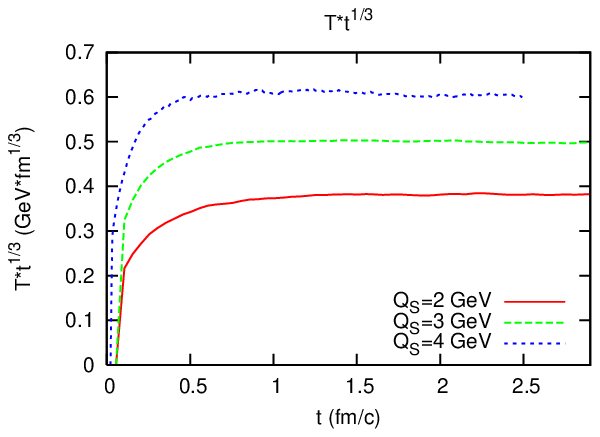}
\end{minipage}
\begin{minipage}[]{50mm}
\epsfxsize=5.0cm
\epsfysize=5.0cm
\epsfbox{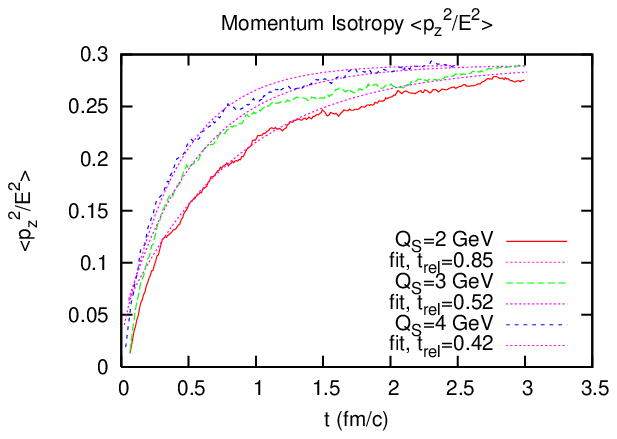}

\end{minipage}
\caption{Time evolution of $\frac{dN}{d\eta}$ (left), of the effective temperature (middle) and
of the momentum anisotropy (right).}
\vskip -0.2cm
\end{figure}

\begin{figure} 
\hskip -0.0cm
\epsfxsize=8.0cm
\epsfysize=4.0cm
\centerline{\epsfbox{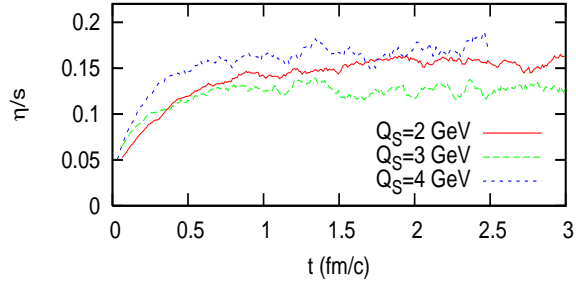}}
\caption{Ratio of the shear viscosity to entropy density ($\alpha_s=0.3$).  }
\label{etaovers}
\vskip -0.2cm
\end{figure}

For all energies a nearly ideal hydrodynamical behavior is observed after $0.5~fm/c$ 
(middle Fig.1). 
The thermalization time lies in the same range when looking at the momentum isotropy.
It is of crucial importance to extract out of these simulations the transport
properties of QCD matter to quantify the dissipative properties of the fluid.
Using standard dissipative hydrodynamics in expanding geometry shear viscosity and ratio $\frac{\eta}{s}$ can be calculated \cite{El:2006xj}: $\eta=\frac{\tau}{4}\left(T_{xx}+T_{yy}-2\cdot T_{zz}\right)$ and 
$s=4n-n\cdot ln\left(\lambda\right)$, where $\lambda $ denotes the gluon fugacity. 
As depicted in Fig. 2, the value $\frac{\eta}{s}\approx 0.15$ proves to be a universal number
within the BAMPS simulations, being nearly independent of $Q_S$.
This is in line also with full 3-dim calculations employing minijets and Glauber geometry
for the initial condition \cite{Xu-next}.  $\frac{\eta}{s}$ basically only depends on the 
employed coupling strength $\alpha_s$ (taken to be 0.3 as default setting).
Hence, within BAMPS, we do not expect any change in the shear viscosity ratio
$\eta /s $ when going from RHIC to LHC.

\end{document}